\documentclass[11pt]{article}
\usepackage[dvips]{graphicx}
\usepackage{amsmath, amssymb}

\newcommand{\vct}[1]{\boldsymbol{#1}}
\newcommand{\mtrix}[3]{\langle \,#1\,|\,#2\,|\,#3\,\rangle}
\newcommand{\avr}[1]{\langle \,#1\,\rangle}

\begin{document}

\baselineskip=14pt

{
\noindent
\Large\textbf{%
The n-th order Moment of the Nuclear Charge Density\\[6pt]
and Contribution from the Neutrons}
}

\vspace{\baselineskip}

\noindent
Haruki Kurasawa\footnote[1]{kurasawa@faculty.chiba-u.jp}$^{1}$ and Toshio Suzuki$^{2}$

\vspace{0.5\baselineskip}

\noindent
$^{1}$\ \parbox[t]{14cm}{Department of Physics, Graduate School
of Science, Chiba University,\\[2pt]
Chiba 263-8522, Japan
}

\vspace{0.5\baselineskip}

\noindent
$^{2}$\ \parbox[t]{14cm}{Research Center for Electron Photon Science, Tohoku University,\\[2pt]
Sendai 982-0826, Japan
}

\begin{center}
\parbox[t]{13cm}{
\small

\baselineskip=12pt

\noindent
The relativistic expression for the $n$-th order moment
of the nuclear charge density is presented.
For the mean square radius(msr) of the nuclear charge density,
the non-relativistic expression, which is equivalent to the relativistic one,
is also derived consistently up to  $1/M^2$ with use of
the Foldy-Wouthuysen transformation. The difference between
the relativistic and non-relativistic expressions for the msr of the
point proton density is also discussed.
The $n(\,\ge 4)$-th order moment of the nuclear charge density
depends on the point neutron density.
The 4-th order moment yields a useful information on the msr
of the point neutron density, and is expected to play an important
role in electron scattering off neutron-rich nuclei. 
}
\end{center}

\vspace{0.5\baselineskip}

\section{Introduction}

At present it is unavoidable for nuclear models to include
unknown parameters. In order to fix their values, some
experimental values have to be used as inputs. One of them is
the mean square radius(msr) of nuclei which is a fundamental 
quantity in nuclear physics.
In most of the previous papers\cite{nlsh,nl3,sly4}, the msr is calculated
using their point proton density, or the one convoluted with a single proton
density, and is compared with the values obtained from the charge
distribution assumed so as to reproduce experimental cross section of
electron scattering.

Electron scattering is an unambiguous tool to examine the nuclear
charge distribution, since the electromagnetic interaction is well
understood theoretically\cite{deforest, oxford}.
Indeed, its msr is determined with high accuracy throughout the periodic
table. For example, the root msr of $^{208}$Pb is reported to be
5.5012(0.0013) fm \cite{frois,newdata}.   

Unfortunately, however, the point proton
density obtained in nuclear models is not determined uniquely from
the experiment. The nuclear charge density deduced from electron
scattering is composed of several elements.
At least in the charge density responsible for
a small momentum transfer $q < 1$ fm$^{-1}$\cite{blunden}, 
they are the point neutron density, the proton and neutron spin-orbit
densities and a single-proton and -neutron
charge density, in addition to the point proton density. 
Their contributions are not distinguished from each other experimentally
or model-independently.

Nuclear models so far have not taken particular care to the
elements other than the point proton density, since their contributions
are believed to be small, and it was not a main purpose of phenomenological
models to reproduce the msr within a few \% accuracy.
Recently, however, more ambitious attempt called {\it ab initio}
calculations has been carried out, trying to explain accurately
the experimental values of the msr\cite{ab,gh}.
Moreover, there are plans to perform experiments to compare
the cross section to the model ones with high accuracy,
aiming to determine a small difference 0.03 $\sim$ 0.05 fm
between the root mean square radii of point-proton and -neutron
distributions\cite{thi}.
In order to respond to these challenges, it is necessary for nuclear models
to make clear a role of each element in the nuclear charge density.
Although main contribution to the charge density comes from the point
proton density, it is not obvious whether or not other elements are
always negligible.
One of the purposes of the present paper is to explore the contribution
of each element in the nuclear charge density to the msr.

Another purpose is to propose a complementary method to the previous
ones\cite{thi} for investigating the neutron density of nuclei.
It is one of the fundamental problems in nuclear physics
how the protons and neutrons are distributed in nuclei.
In contrast to the proton distribution, however,
the neutron distribution is not well known yet, since there is no
simple way to explore it experimentally.
For example, in elastic electron scattering, 
the cross section is strongly dominated by the proton charge
distribution, and the contribution from the neutrons
is hidden behind the one from protons. As a result, it is hard to
extract information on the neutron density by the analysis of
the charge density profile.
In hadron scattering\cite{pp}, there is a different kind of difficulties.
Although contributions to the cross section
from neutrons and protons are comparable, they are not distinguishable from
each other, because in the strong interaction, the reaction mechanism
is not well understood, and the physical meaning of the parameters employed
in the analyses is not obvious\cite{thi}.  
As a unique experiment to observe directly the neutron weak charge
density, the measurement of the parity-violating asymmetry in the
polarized-electron scattering has recently been performed\cite{abra,cjh}.
It is a promising, but very difficult and time-consuming experiment.
At present, the value of the
form factor is available only for $^{208}$Pb at a single value
of the momentum transfer, $q = 0.475\,\textrm{fm}^{-1}$,
with the error of about 10\%. 
Thus, it does not seem that the neutron density profiles are extracted soon
from experiment without the help of nuclear models.
In the present paper, it will be shown that 
instead of discussing the charge density profiles themselves,
the analyses of their moments provide us with the useful information on
the neutron distribution in nuclei.

In the following section, the relativistic charge density will be defined.
In \S\ref{nm}, the expression of the $n$-th order moment will be derived. 
In \S\ref{s2nd} and \S\ref{s4th}, the 2nd(msr) and 4th order moments will
be discussed in detail, respectively.
For the msr, the non-relativistic expression, which is equivalent
to the relativistic one up to order $1/M^2$, will be derived consistently,
according to the  Foldy-Wouthuysen(F-W) transformation. 
The new term which has not been discussed so far is obtained as
a relativistic correction. 
The difference between the mean square radii of the relativistic
and non-relativistic point proton densities will also be shown
in an analytic way.
Unlike the msr, the $n(\,\ge 4)$-th order moment depends
on the point neutron density.
In the 4th order moment, the msr of the point neutron density
and the 4th order moments of the neutron spin-orbit density
play a crucial role, as corrections to the main term from the point
proton density.

The present results may be useful for reducing ambiguity of nuclear model
parameters, and hence, for recent detailed discussions of
the nuclear proton and neutron density profiles\cite{thi}.
Moreover, there is a possibility that not only the change of the
proton distribution, but also the one of the neutron distribution
will be explored throughout neutron-rich and proton-rich nuclei
with the conventional and well-established electron scattering
experiment\cite{tsukada, tsuda}.

The final section is devoted to a brief summary of the present paper.

\section{Relativistic nuclear charge density}

The relativistic charge density of the nuclear ground state $|\,0\,\rangle$
is given by\cite{ks}
\begin{equation}
 \rho_c(r) = \int \frac{d^3q}{(2\pi)^3}\exp(-i\vct{q}\!\cdot\!\vct{r})
 \tilde{\rho}(q).\label{rcd}
\end{equation}
Its Fourier component is described as
\begin{equation}
\tilde{\rho}(q) = \mtrix{0}{\sum_{k=1}^A
\exp(i\vct{q}\!\cdot\!\vct{r_k})\left(F_{1k}(q^2)+\frac{\mu_k}{2M}F_{2k}(q^2)
\vct{q}\!\cdot\!\vct{\gamma}_k\right)}{0},
\end{equation}
where $A$ denotes the nucleon number of the system, $F_1$
and $F_2$ the Dirac and Pauli form factor of the nucleon, respectively,
$\mu_k$ the anomalous magnetic moment, and $M$ the nucleon mass\cite{bd}.
Throughout the present paper, the following values will be used,
$\mu_k=1.793$ and $-1.913$ for the proton(p) and neutron(n), respectively
and $M=939$ MeV.
The center-of-mass corrections will not be taken into account. 

The above matrix element is rewritten in terms of Sachs form factor, 
\begin{equation}
\tilde{\rho}(q) =\int d^3x\exp(i\vct{q}\!\cdot\!\vct{x})
 \sum_\tau\Bigl(G_{E\tau}(q^2)\rho_\tau(x)+F_{2\tau}(q^2)W_\tau(x)\Bigr),
 \label{fourier}
\end{equation}
where $\tau$ represents p and n. The Sachs form factor
is related to the Dirac and Pauli form factor as\cite{bd}
\begin{equation}
G_{E\tau}(q^2)=F_{1\tau}(q^2)-\mu_\tau q^2F_{2\tau}(q^2)/(4M^2).\label{sachs}
\end{equation}
The point nucleon density $\rho_\tau$ and the spin-orbit density $W_\tau$
are given by\cite{ks}
\begin{align}
\rho_\tau(r)&= \mtrix{0}{\sum_{k\in\tau} \delta(\vct{r}-\vct{r}_k)}{0},\\
W_\tau(r)&= \frac{\mu_\tau}{2M}\left(-\frac{1}{2M}\vct{\nabla}^2\rho_\tau(r)
+i\vct{\nabla}\!\cdot\! \mtrix{0}{\sum_{k\in\tau}
\delta(\vct{r}-\vct{r}_k)\vct{\gamma}_k}{0}\right).
\end{align}
The first equation satisfies $\int d^3r \, \rho_\tau (r) = Z$ for $\tau = p$
and $N$ for $\tau = n$, respectively, while the last equation 
$\int d^3r\, W_\tau(r)=0$, as should be. Their explicit forms
in relativistic  nuclear mean field models are written as\cite{ks}
\begin{align} 
\rho_\tau(r)&= \sum_{\alpha\in\tau} \frac{2j_\alpha+1}{4\pi r^2}
 \left(G_\alpha^2 + F_\alpha^2\right),\\
W_\tau(r)&= \frac{\mu_\tau}{M}\sum_{\alpha\in\tau} \frac{2j_\alpha+1}{4\pi r^2}
 \frac{d}{dr}\left(\frac{M-M^*(r)}{M}G_\alpha F_\alpha
+ \frac{\kappa_\alpha +1}{2Mr}G_\alpha^2 -
\frac{\kappa_\alpha - 1}{2Mr}F_\alpha^2\right),
\label{so}
\end{align}
In the above equations, $j_\alpha$ denotes the total angular momentum
of a single-particle, $\kappa_\alpha=(-1)^{j_\alpha-\ell_\alpha
 +1/2}(j_\alpha+1/2)$,
$\ell_\alpha$ being the orbital angular momentum, and 
$G_\alpha(r)$ and $F_\alpha(r)$ stand for the radial parts of
the large and small component of the single-particle wave function,
respectively, with the normalization,
\begin{equation}
\int_0^\infty\! dr \left(G_\alpha^2 + F_\alpha^2\right)=1.\label{norm}
\end{equation}
In Eq.(\ref{so}), the effective nucleon mass is defined
by $M^*(r)=M+V_\sigma(r)$, $V_\sigma(r)$ being the $\sigma$
meson-exchange potential which behaves in the same way
as the nucleon mass in the equation of motion.
The spin-orbit density is a relativistic correction due to the anomalous
magnetic moment of the nucleon, and its role is enhanced by
the effective mass in relativistic nuclear models\cite{ks}.
The reason why Eq.(\ref{so}) is called the spin-orbit density
will be seen later in discussing its non-relativistic reduction.

The relativistic nuclear charge density Eq.(\ref{rcd}) is finally
written as,
\begin{equation}
 \rho_c(r) =\sum_\tau\Bigl( \rho_{c\tau}(r) + W_{c\tau}(r) \Bigr)
\end{equation}
by convoluting a single-proton and -neutron density, 
\begin{align}
\rho_{c\tau}(r)&= 
\frac{1}{r}\int_0^\infty dx\, x\rho_\tau(x)\Bigl( g_\tau(|r-x|)
-g_\tau(r+x)\Bigr), \\[4pt]
W_{c\tau}(r)&= \frac{1}{r}\int_0^\infty dx\,x W_\tau(x)
\Bigl( f_{2\tau}(|r-x|)-f_{2\tau}(r+x)\Bigr),
\end{align}
with the functions,
\begin{equation}
g_\tau(x)= \frac{1}{2\pi}\int_{-\infty}^\infty dq\, e^{iqx}G_{E\tau}(q^2),
 \hspace{1cm}
f_{2\tau}(x)= \frac{1}{2\pi}\int_{-\infty}^\infty dq\, e^{iqx}F_{2\tau}(q^2).
\end{equation}
The momentum-transfer dependence of the nucleon form factors
is not well known yet, and are still under discussions
theoretically\cite{sachs,licht,gmiller}.
Experimentally also there are various versions to fit
the electron scattering data at present\cite{kelly,kelly2}.
In this paper, the following Sachs and Pauli form factors 
will be employed\cite{ks,bertozzi,eden,Mey,plat},
\begin{align}
G_{Ep}(q^2)&= \frac{1}{(1+r_p^2q^2/12)^2},
 \qquad F_{2p}=\frac{G_{Ep}(q^2)}
 {1+q^2/4M^2}, \label{expff}\\[4pt]
G_{En}(q^2)&= \frac{1}{(1+r_+^2q^2/12)^2}- \frac{1}{(1+r_-^2q^2/12)^2},\qquad
 F_{2n}=\frac{G_{Ep}(q^2)-G_{En}(q^2)/\mu_n}{1+q^2/4M^2},\nonumber
\end{align}
with
\[
 r_p=0.81\, \textrm{fm}, \qquad r_{\pm}^2=(0.9)^2
  \mp0.06 \, \textrm{fm}^2.
  \nonumber
\]
These form factors have been determined by fitting electron
scattering data within a relativistic framework, but we note that
there are still discussions
on the values of $r_p$ and $ r_{\pm}^2$ themselves\cite{sick}.

\section{Relativistic expression of the n-th order moment}\label{nm}

The relativistic charge density Eq.(\ref{rcd}) satisfies
$\int d^3r \, \rho_c(r) = Z$.
The mean $2n$-th order moment $\avr{r^{2n}}_c$ of the nuclear
charge distribution is given by
\begin{equation}
\avr{r^{2n}}_c =\sum_\tau \avr{r^{2n}}_{c\tau},
\qquad
 Z\avr{r^{2n}}_{c\tau}=\int d^3r\, r^{2n}\rho_{c\tau}(r)
  =(-\vct{\nabla}_q^2)^n
  \tilde{\rho}_\tau(q)|_{q=0}.\label{nth}
\end{equation}
Here, according to Eq.(\ref{fourier}), we have defined the notations  
\[
\tilde{\rho}_\tau(q) 
= \int d^3r\, e^{i\vct{q}\cdot\vct{r}}C_\tau(q^2, r)
=\int d^3r\, j_0(qr)C_\tau(q^2, r),
\]
with
\[
 C_\tau(q^2, r)=G_{E\tau}(q^2)\rho_\tau(r) +F_{2\tau}(q^2)W_\tau(r).
\]
In order to calculate Eq.(\ref{nth}), it is convenient to define
the integral,
\begin{equation}
\tilde{F}(q)=\int d^3r\, j_0(qr)G(q^2)F(r),\qquad
G(x)=\frac{1}{(1+\Lambda x)^{\alpha+1}},\qquad (\,\alpha = 0, 1, \cdots\,),
\end{equation}
since the Pauli form factor is written in the form,
\begin{align*}
 F_{2p}
&=\frac{1}{(a-b)^2}\left(-\frac{ab}{1+aq^2}+\frac{a(a-b)}{(1+aq^2)^2}
 +\frac{b^2}{1+bq^2}\right), \\[4pt]
F_{2n}
&=F_{2p}-\frac{1}{\mu_n}
\Bigl( F_{2p}(r_p\rightarrow r_+)-F_{2p}(r_p \rightarrow r_-)\Bigr),
\end{align*}
with use of the notations $a=r_p^2/12$ and $b=1/4M^2$.
Then, we have
\begin{equation}
(-\vct{\nabla}_q^2)^n\tilde{F}(q)=\sum_{k=0}^{2n}
{}_{2n}\mathrm{C}_k
\int d^3r\, F(r)J_k,
\end{equation}
where $_n\mathrm{C}_k$ denotes the binomial coefficient and 
\begin{align*}
 J_{2k}&= (-1)^kr^{2n-2k}\frac{\sin(qr)}{qr}\sum_{m=0}^{k}\frac{(2k)!G^{(2k-m)}
(q^2)}{m!(2k-2m)!}(2q)^{2k-2m}, \quad (0\le k\le n), \\[4pt]
 J_{2k-1}&= (-1)^kr^{2n-2k+1}\frac{\cos(qr)}{qr}\sum_{m=0}^{k-1}
  \frac{(2k-1)!G^{(2k-1-m)}(q^2)}{m!(2k-1-2m)!}(2q)^{2k-1-2m}, \quad
  (1\le k \le n),
\end{align*}
with
\[
G^{(n)}(x)=\frac{(\alpha+n)!}{\alpha !}\frac{(-\Lambda)^n}
 {(1+\Lambda x)^{\alpha+1+n}}.
\]
In the limit $q \rightarrow 0$, $J_{2k}$ and $J_{2k-1}$ become of
\[
J_{2k},\,J_{2k-1}\longrightarrow (2k)!\,{}_{\alpha+k}\mathrm{C}_\alpha
 r^{2n-2k} \Lambda^k,
\]
which yields
\begin{equation}
(-\vct{\nabla}_q^2)^n\tilde{F}(q)|_{q=0}=\int d^3r\,r^{2n}F(r)
+\sum_{k=1}^n\frac{(2n+1)!}{(2n+1-2k)!}\,
{}_{\alpha+k}\mathrm{C}_\alpha\Lambda^k
\int d^3r\, r^{2n-2k}F(r).
\end{equation}
In the present paper, we will discuss in detail
the case $n=1$, and $2$, which are
given explicitly as,
\begin{align}
(-\vct{\nabla}_q^2)\tilde{F}(q)|_{q=0}
&= \int d^3r\,r^2F(r)
+6(\alpha+1)\Lambda\int d^3r\,F(r),  \label{2nd}\\[4pt]
(-\vct{\nabla}_q^2)^2\tilde{F}(q)|_{q=0}
&= \int d^3r\,r^4F(r)
+20(\alpha+1)\Lambda\int d^3r\,r^2F(r) \nonumber\\ 
&\hphantom{=\ \int d^3r\, r^4F(r)}
   +60(\alpha+1)(\alpha+2)\Lambda^2
\int d^3r\, F(r).\label{4th}
\end{align}

\subsection{The 2nd order moment of the charge density}\label{s2nd}

Eq.(\ref{nth}) and Eq.(\ref{2nd}) provide us with the
relativistic expression for the msr of nuclear charge
 distribution\cite{horowitz}, 
\begin{equation}
\avr{r^2}_c
= 
\avr{r^2}_p  + r_p^2 +(r_+^2-r_-^2)\frac{N}{Z}
+\avr{r^2}_{W_p}+\avr{r^2}_{W_n}\frac{N}{Z},
\label{msr}
\end{equation}
where we have defined with $N_p=Z$ and $N_n=N$,
\[
\avr{r^k}_\tau = \frac{1}{N_\tau}\int d^3r\, r^k\rho_\tau(r),
\qquad
\avr{r^k}_{W_\tau}=\frac{1}{N_\tau}\int d^3r\, r^kW_\tau(r).
\]
The msr depends on a single-proton and a single-neutron size
through the second and the third term of the above equation.
They are not in a negligible order, but unfortunately, it seems that
their experimental values are still
under discussion \cite{sick}.
We note, however, that their contributions to the msr is almost
eliminated in taking
the difference between the values for two nuclei, for example,
in discussing isotope and isotone shift, or mirror nuclei.
Then, the msr is approximately given by the point proton radius
and the proton and neutron spin-orbit densities.  Among them, if protons fill
the shells of spin-orbit partners, contributions from their spin-orbit term
is negligible, as will be shown later. In that case, the msr of the charge
density is given by the ones of the point proton density and
the neutron spin-orbit density in relativistic models.

In non-relativistic models, the following expression of the msr
has been widely used\cite{ong},
 \begin{equation}
\avr{r^2}_{c,\mathrm{nonrel}} 
=\avr{r^2}_{p,\mathrm{nonrel}} + r_p^2 
+(r_+^2-r_-^2) \frac{N}{Z}
-\frac{1}{Z}\sum_{i=1}^A\frac{\mu_i}{M^2}(\kappa_i+1)+\frac{3}{4M^2}.
 \label{nmsr}
 \end{equation}
In the r.h.s. of the above equation, the first term stands for the msr of
the point proton density calculated with non-relativistic wave functions.
The values of the second and the
third term are taken from the Sachs form factors determined experimentally
within a relativistic framework, as used in Eq.(\ref{msr}).
The 4th term is added as a non-relativistic reduction of the spin-orbit
terms in Eq.(\ref{msr})\cite{bertozzi,ong},
and the 5th one is explained as coming
from the Darwin-Foldy(D-F) term which is a relativistic correction to
the nuclear charge density\cite{friar}. In fact, however, the consistency
of these terms is doubtful,
since they should be calculated consistently, according to
the Foldy-Wouthuysen(F-W) unitary transformation of the four-component
framework to the two-component one\cite{oxford,bd,messiah}.

The F-W transformation for Dirac equation with electromagnetic field
has been performed by various authors\cite{bertozzi,macvoy,nishizaki}.
For example, Nishizaki and the present authors\cite{nishizaki} have  obtained
a single-particle charge operator $\hat{\rho}(q)$ up to order $1/M^2$,
\begin{align}
\hat{\rho}(q) = e^{i\vct{q}\cdot\vct{r}}\Bigl( D_1(q^2)+ iD_2(q^2)
\vct{q}\!\cdot\!\vct{p}_\sigma \Bigr),\qquad
(\,\vct{p}_\sigma=\vct{p}\times\vct{\sigma}\,),
\label{fw}
\end{align}
where $D_1$ and $D_2$ are defined as
\begin{align}
 D_1(q^2)&= F_{1\tau}-\frac{q^2}{8M^2}\left(F_{1\tau}+2\mu_\tau F_{2\tau}
 \right), \label{d0}\\[4pt]
 D_2(q^2)&= \frac{F_{1\tau}+2\mu_\tau F_{2\tau}}{4M^2}.\label{d2}
\end{align}
The relativistic corrections are those proportional to $1/M^2$.  
The second term  of $D_1$ is called Darwin-Foldy term,
while $D_2$ the spin-orbit term in ref.\cite{macvoy}.
The first term and the $F_2$ part of the Darwin-Foldy term in $D_1$
are replaced by the Sachs form factor, according to Eq.(\ref{sachs}),
\begin{equation}
D_1(q^2)=G_{E_\tau} - \frac{q^2}{8M^2}F_{1\tau}.\label{d1}
 \end{equation}
Then the second derivative of Eq.(\ref{fw}) yields
\begin{equation}
 -\vct{\nabla}_{\vct{q}}^2\hat{\rho}(q)|_{\vct{q}=0}
=D_1(0)r^2+2D_2(0)\vct{\ell}\!\cdot\!\vct{\sigma}-6D_1'(0),\label{derivative}
\end{equation}
where $D_1'$ denotes the first derivative with respect to $q^2$.
Finally, Eq.(\ref{derivative}) 
provides us with the expression of the msr up to $1/M^2$,
\begin{equation}
\avr{r^2}_c
 =\frac{1}{Z}\mtrix{0}{\sum_{k=1}^Z r_k^2}{0}_{\mathrm{nonrel}}
 -6G'_{E_p}(0)-6G'_{E_n}(0)\frac{N}{Z}+R, \label{correct}
\end{equation}
where $R$ represents the contributions of order $1/M^2$ 
except for the one from the $F_2$ part included in the Sachs form factor
of Eq.(\ref{d1}),
\begin{equation}
R= \frac{1}{M^2}\left(\frac{1}{Z}\sum_{k=1}^A\mu_k
 \mtrix{0}{\vct{\ell}_k\!\cdot\!\vct{\sigma}_k}{0}_{\mathrm{nonrel}}+\frac{3}{4}
 +\frac{1}{2Z}\sum_{k=1}^Z\mtrix{0}
 {\vct{\ell}_k\!\cdot\!\vct{\sigma}_k}{0}_{\mathrm{nonrel}} \right).\label{r}
\end{equation}
The terms of the r.h.s. in Eq.(\ref{correct}) are formally consistent
with each other, but in the present framework the values of
$G'_{E_\tau}(0)$ up to order $1/M^2$ are unknown.
In the relativistic expression Eq.(\ref{msr}), they are taken
from Eq.(\ref{expff}) determined by experiment with use of the relationship,
\begin{equation}
 6G'_{E_p}(0)=-r_p^2\, , \hspace{5mm}   6G'_{E_n}(0)=-(r_+^2-r_-^2)\,.
\label{pn}
\end{equation}
If the same values are employed as in Eq.(\ref{nmsr}) ,
the consistency in Eq.(\ref{correct}) becomes obscure.
Since the structure of $G_{E_\tau}$ is not well known theoretically
at present\cite{sachs,licht,gmiller}, this ambiguity remains
in the expression of Eq.(\ref{correct}).

In comparing Eq.(\ref{correct}) with Eq.(\ref{nmsr}), the former
equation has an additional term in $R$, which is the last one in Eq.(\ref{r}).
It stems from the $F_{1p}$ part of $D_2$ in Eq.(\ref{d2}),
while the first term in $R$ from its $F_2$ part.
The last term contributes additively to the proton part of
the first one, cancelling more its neutron part with the negative sign
of $\mu_k$. 

We note that the second term $3/4M^2=0.0331\,\textrm{fm}^{2}$ in $R$
comes from the $F_{1p}$ part of the Darwin-Foldy term in Eq.(\ref{d0}),
whose $F_2$ part is taken in $G_{E_\tau}$ of Eq.(\ref{d1}).

When the nuclear part of the hamiltonian is different from the
Dirac equation, the F-W transformation yields different relativistic
corrections.
For the mean field hamiltonian in the $\sigma$-$\omega$
model, Nishizaki et al.\cite{nishizaki} have obtained, instead of
Eqs.(\ref{d0}) and (\ref{d2}),
\begin{align}
 D_1(q^2)&= F_{1\tau}-\frac{q^2}{8M^{*2}(r)}\left(F_{1\tau}
+2\mu_\tau F_{2\tau}\frac{M^*(r)}{M}
 \right), \label{d0m}\\[4pt]
 D_2(q^2)&= \frac{1}{4M^{*2}(r)}\left(F_{1\tau}+2\mu_\tau F_{2\tau}\frac{M^*(r)}
 {M}\right).\label{d2m}
\end{align}
In this case, Eq.(\ref{r}) is replaced by
\begin{equation}
R= \frac{1}{MM^*Z}\sum_{k=1}^A\mu_k
 \mtrix{0}{\vct{\ell}_k\!\cdot\!\vct{\sigma}_k}{0}_{\mathrm{nonrel}}
+\frac{1}{M^{*2}}\left(\frac{3}{4}
 +\frac{1}{2Z}\sum_{k=1}^Z\mtrix{0}
 {\vct{\ell}_k\!\cdot\!\vct{\sigma}_k}{0}_{\mathrm{nonrel}}\right),\label{rr}
\end{equation}
in the approximation $M^*(r)\approx M^*$.
The second term is proportional to $1/M^{*2}$, while the first term to
$1/MM^*$, because of the definition of $F_{2\tau}$.   
Most of the relativistic models\cite{nl3} has $M^*\approx 0.6M$ 
which enhances the relativistic corrections.
In particular, the D-F correction $3/4M^2$ is enhanced by $(M/M^*)^2$,
yielding $3/4M^{*2} \approx 0.0920\,\textrm{fm}^{2}$.
The enhancement of the $F_2$ part in $D_1$, Eq.(\ref{d0m}),
is absorbed into the Sachs form factor, if it is taken from experiment.

It may be instructive to show in a more naive way where the last two terms
in $R$ of Eq.(\ref{r}) stem from in the non-relativistic reduction,
and why the D-F corrections are enhanced by the effective mass
as in Eq.(\ref{rr}) .
Since they are independent of $\mu_k$,
it is expected that they are contained in the first term of
the relativistic expression Eq.(\ref{msr}).

In using the Dirac equation\cite{sw}, the $n$-th order moment
of the density distribution as to the small component $F(r)$ is expressed
in terms of the large component $G(r)$,
\begin{equation}
\avr{r^n}_F =\int_0^\infty dr\, r^n F^2(r)
 =\frac{1}{(E+M)^2}\int_0^\infty\! dr
 \left(\frac{n(n-1-2\kappa)}{2}r^{n-2}G^2+(E^2-M^2)r^nG^2 \right),
 \nonumber
\end{equation}
which provides us with
\begin{equation}
\avr{r^2}_F = \frac{1-2\kappa}{(E+M)^2}+\frac{E-M}{E+M}
 \avr{r^2}_G\,,\qquad 
\avr{r^2}_G=\int_0^\infty dr\, r^2 G^2(r).
 \end{equation}
In the first term of the r.h.s., we have used the approximation
\begin{equation}
\int_0^\infty dr\,G^2(r)=1\, ,\nonumber
 \end{equation}
corresponding to the approximation for the small component,
\begin{equation}
\int_0^\infty dr\, F^2(r) = \frac{E-M}{E+M}\,.\nonumber
\end{equation}
Hence, the msr of the single-particle is written as
\begin{equation}
 \avr{r^2} = \avr{r^2}_G + \avr{r^2}_F
 \approx  \avr{r^2}_G + 
\frac{3+2\avr{\vct{\ell}\!\cdot\!\vct{\sigma}}_G}{4M^2},\label{ls}
 \end{equation}
where we have used the fact that
\begin{equation}
\avr{\vct{\ell}\!\cdot\!\vct{\sigma}}_G =-(1+\kappa)\, .\label{lsk}
\end{equation}
The second term of the r.h.s in Eq.(\ref{ls}) is nothing but
the last two terms in $R$. In this way, they are understood as a contribution
of the small component to the msr.

In relativistic mean field models, the nuclear part of the hamiltonian
contains one-body potentials.
In the $\sigma-\omega$ model, the equations for the radial parts
of the wave function are written as\cite{sw}
\begin{align}
\frac{dG}{dr}&= -\frac{\kappa}{r}G(r)+\frac{F(r)}{\lambda(r)}\, ,
 \qquad \lambda(r)=\frac{1}{E+M+V_\sigma(r)-V_0(r)}\, ,\\[4pt]
\frac{dF}{dr}&= \frac{\kappa}{r}F(r)-\varepsilon(r)G(r)\, ,
 \qquad \varepsilon(r)=E-M-V_\sigma(r)-V_0(r)\, ,
 \end{align}
where $V_0(r)$ denotes the potential due to
the $\omega$-meson exchange.
In this case, we obtain
\begin{equation}
 \avr{r^n}_F = \int_0^\infty dr\left(\frac{n(n-1-2\kappa)}{2}
\lambda^2 r^{n-2}G^2+\frac{d \lambda}{dr} \Bigl( n\lambda r^{n-1}G^2-r^nFG \Bigr)
+\varepsilon\lambda r^nG^2\right), \nonumber 
\end{equation}
which gives
\begin{equation}
\avr{r^2} \approx \avr{r^2}_G 
+ \bigl(3+2\avr{\vct{\ell}\!\cdot\!\vct{\sigma}}_G \bigr)
\int_0^\infty dr\, \lambda^2G^2.\label{correc}
 \end{equation}
This is not exactly the same as, but similar to the corresponding term
of Eq.(\ref{rr}), because of $V_\sigma(r) \approx -V_0(r)$ in the
relativistic models. Thus, it is understood why the last two terms
of Eq.(\ref{rr}) are enhanced by the nucleon effective mass.

The above derivation of the relativistic corrections implies that,
when we compare the relativistic and  non-relativistic results with
each other, the difference between $\avr{r^2}_p$ in Eq.(\ref{msr})
and $\avr{r^2}_{p,\mathrm{nonrel}}$ in Eq.(\ref{correct})
should be taken into account.
If $\avr{r^2}_{p,\mathrm{nonrel}}$ is fixed so as to reproduce
experimental values of the msr,
$\avr{r^2}_{p,\mathrm{nonrel}}$ may contain a part of
the relativistic correction in Eq.(\ref{correc}).

The 4th term of Eq.(\ref{nmsr}) is the same as the first term in Eq.(\ref{r}).
In fact, it is derived by various ways\cite{bertozzi,ong} without concerning
the consistency with other terms. 
For example, if we neglect simply the small component $F_\alpha$
in the relativistic spin-orbit density Eq.(\ref{so}), we have
\begin{equation}
W_\tau(r)\rightarrow \frac{\mu_\tau}{2M^2r^2}\frac{d}{dr}r\sum_\alpha
 \frac{2j_\alpha+1}{4\pi r^2}(\kappa_\alpha+1)G_\alpha^2,\label{nso}
\end{equation}
which yields
\begin{equation}
\avr{r^k}_{W_\tau} \approx -\frac{1}{N_\tau}\frac{k\mu_\tau}{2M^2}
\sum_{\alpha \,\epsilon\,\tau}(2j_\alpha+1)(\kappa_\alpha+1)
\int_0^\infty dr\, r^{k-2}G_\alpha^2(r).\label{nsom}
\end{equation}
In the case of $k=2$, together with  Eq.(\ref{norm}) neglecting $F_\alpha$,
the above equation gives the spin-orbit correction of Eq.(\ref{nmsr})
and Eq.(\ref{r}), but does not the one in Eq.(\ref{rr})
from the F-W transformation.

We note that Eq.(\ref{nso}) is written as\cite{ks},
\begin{equation}
W_\tau(r)\rightarrow -\frac{\mu_\tau}{2M^2r^2}\frac{d}{dr}r
\mtrix{0}{\sum_k\delta(\vct{r}-\vct{r}_k)\vct{\sigma}_k
\!\cdot\!\vct{\ell}_k}{0}.
\end{equation}
The above is the reason why $W_\tau(r)$ is called
the spin-orbit density\cite{ks}.
The previous authors\cite{bertozzi} have pointed out that there is
the exact cancellation of the spin-orbit terms between the fully
occupied spin-orbit partners in their non-relativistic limit.
This fact is easily understood from the above expression, and also valid
in the last term of Eq.(\ref{r}).

Now, we investigate the msr of the nuclear charge distribution  given
by Eq.(\ref{msr}) in the relativistic models and by Eq.(\ref{correct})
in the non-relativistic framework.
Table 1 shows how each term contributes to the msr
in the cases of $^{40}$Ca, $^{48}$Ca, and $^{208}$Pb.
Among the experimental data available at present\cite{newdata,sudadata},
those for the above three nuclei are suitable as examples for the present
purpose, since they have been well investigated using a mean field
approximation, and used to fix the parameters of the model hamiltonian.
In this approximation, $^{48}$Ca is described as the $f_{7/2}$ closed
shell nucleus, while  $^{40}$Ca is the double closed shell one,
and in $^{208}$Pb, protons occupy up to $h_{11/2}$ shell and neutrons
up to $i_{13/2}$ shell.

\begin{table}
\begin{tabular}{|l||r|r|r|r|r||r|} \hline
\rule{0pt}{12pt} & 
 $\avr{r^2}_{p}+r_p^2$  &
 $(r_+^2 - r_-^2)N/Z$ &
 $\avr{r^2}_{W_p}$ &
 $\avr{r^2}_{W_n}N/Z$  &
\multicolumn{1}{c||}{$\avr{r^2}_{c}$} &
\multicolumn{1}{c|}{Exp.} \\ \hline
\rule{0pt}{12pt}%
NL3        &          &           &          &           &          &          \\
$^{40}$Ca  & $12.060$ & $-0.1200$ & $0.0222$ & $-0.0244$ & $11.938$ & $11.90$ \\
$^{48}$Ca  & $12.073$ & $-0.1680$ & $0.0263$ & $-0.1573$ & $11.774$ & $11.91$ \\
$^{208}$Pb & $30.467$ & $-0.1844$ & $0.1054$ & $-0.1460$ & $30.242$ & $30.28$ \\
\hline
\rule{0pt}{12pt}%
NL-SH      &          &           &          &           &          &          \\
$^{40}$Ca  & $11.928$ & $-0.1200$ & $0.0225$ & $-0.0248$ & $11.806$ & $11.90$ \\
$^{48}$Ca  & $11.998$ & $-0.1680$ & $0.0268$ & $-0.1594$ & $11.698$ & $11.91$ \\
$^{208}$Pb & $30.304$ & $-0.1844$ & $0.1071$ & $-0.1482$ & $30.304$ & $30.28$ \\
\hline
\rule{0pt}{12pt}%
SLy4       &          &           &          &           &          &          \\
$^{40}$Ca  & $12.404$ & $-0.1200$ & $0.0000$ & $ 0.0000$ & $12.318$ & $12.18$ \\
$^{48}$Ca  & $12.602$ & $-0.1680$ & $0.0000$ & $-0.1014$ & $12.366$ & $12.11$ \\ 
$^{208}$Pb & $30.448$ & $-0.1844$ & $0.0579$ & $-0.0865$ & $30.284$ & $30.25$ \\ \hline
\end{tabular}
\caption{
The mean square radius(msr) of the charge distribution
of $^{40}$Ca, $^{48}$Ca and $^{208}$Pb in the unit of fm$^2$.
The calculated values, using parameters of the relativistic nuclear
models NL3\cite{nl3} and NL-SH\cite{nlsh},
and of the non-relativistic one SLy4\cite{sly4}, are listed.
The experimental values are those used in the nuclear models
to fix their parameters. For details, see the text.  
}
\end{table}

In Table 1, the sum of the first two terms and each of the rest
in Eq.({\ref{msr}}) are listed separately, in the relativistic cases,
NL3\cite{nl3} and NL-SH\cite{nlsh}.
In non-relativistic calculations of Eq.(\ref{correct}) with SLy4,
its second and the third term are taken from Eq.(\ref{pn}), as in
the relativistic models.
The values of the first term in the relativistic correction $R$
given by Eq.(\ref{r}) are listed
as $\avr{r^2}_{W_p}$ and $\avr{r^2}_{W_p}N/Z$,
while the second term, $3/4M^2=0.0331\,\textrm{fm}^2$, is included
in $\avr{r^2}_c$ in the Table 1.
The last term of $R$ in Eq.(\ref{r}) does not contribute to 
$\avr{r^2}_c$ in Ca, but does in $^{208}$Pb. Its value,
0.0162 fm$^2$, is added to $\avr{r^2}_c$ of $^{208}$Pb.  

The experimental values of the msr employed as inputs for fixing the parameters
of NL3 and SLy4 are also listed in Table 1,
according to the refs.\cite{nl3,sly4}.
For NL-SH, the same values as the ones of NL3 are putted, which are taken
from ref.\cite{sudadata}. Unfortunately, however,
it is not clear for the authors
which corrections to the msr of the calculated point proton densities
are taken into account in the refs.\cite{nlsh, nl3, sly4},
in reproducing the quoted experimental data.  
 
As seen in the Table 1, the corrections from the last three terms
to the first two terms in Eq.(\ref{msr}) are less than $3\%$
in relativistic models, in spite of the fact that
the spin-orbit densities are enhanced by the effective mass\cite{ks}.
The sum of the first two terms for $^{48}$Ca, however, is slightly 
larger than for $^{40}$Ca, while the total sum value of the former is
smaller than the latter.

In non-relativistic models\cite{sly4}, the ambiguity coming from the
relativistic corrections in Eq.(\ref{correct}) may be within $2\%$.
We note, however, that if non-relativistic model fixes parameters,
so as to reproduce the experimental values with the msr of the point proton
density, it may contain relativistic corrections implicitly.
In fact, Table 1 shows that the msr value of the point proton density
in SLy4 is similar to the ones in the relativistic models
which include a non-negligible relativistic correction enhanced by
the effective mass.  
When one employs experimental values of the msr as inputs in
phenomenological models\cite{nlsh, nl3,sly4},
or when one compares phenomenological
models with each others for precise discussions\cite{thi}, 
these small corrections should be taken into account carefully.
 
Table 1 shows also that the cancellation of the spin-orbit terms
between the spin-orbit partners does not hold exactly 
in the relativistic models, as was already pointed out in ref.\cite{miller}.

\subsection{The 4th order moment of the charge density}\label{s4th}

The function $r^6\rho_c(r)$ has a peak around the nuclear surface
in a similar way as $r^4\rho_c(r)$.
As a result, it is expected that
the 4th order moment also reflects well the structure of the nuclear
surface, as the msr does.
This fact implies that it is useful to investigate the 4th order
moment for understanding more details of the nuclear surface structure.
Compared with the msr, however, the 4th order moment has not been
explored in detail so far.

\begin{figure}[ht]
\centering{\includegraphics{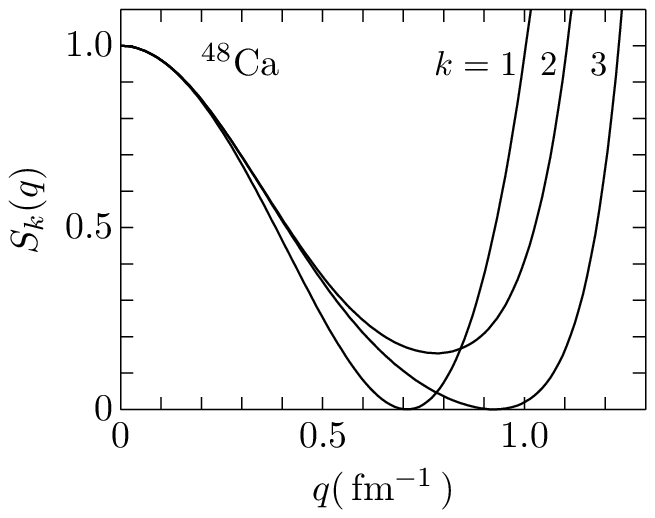}\qquad%
\includegraphics{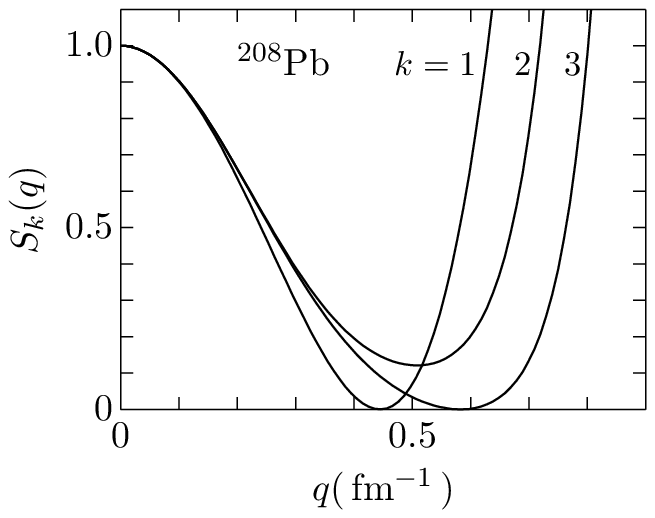}%
}
\caption{
The contribution of the $n$-th order moment of the charge density to
the form factor squared for elastic electron scattering off
$^{48}$Ca and $^{208}$Pb in PWBA\cite{suda}. For details, see the text.
}
\end{figure}

Fig.1 shows the contribution of the $n$-th order moment of the charge
density to the form factor squared for elastic electron scattering off
$^{48}$Ca and $^{208}$Pb\,\cite{suda}, which is defined as
\begin{equation}
S_k(q) = \Bigl|\sum_{n=0}^{k} F_{2n}(q)\Bigr|^2\,, 
\hspace{5mm}F_{2n}(q)=(-1)^n
\frac{q^{2n}\avr{r^{2n}}_c}{(2n+1)!}\,,
\end{equation}
where the $2n$-th order moment is calculated using the Fourier-Bessel analyses
of the experimental data\cite{sudadata}.
It is seen that the msr dominates the form factor in $^{48}$Ca up to
$q\approx 0.3\,\textrm{fm}^{-1}$, and in $^{208}$Pb up to
$q\approx 0.2\,\textrm{fm}^{-1}$.
Above these momentum transfer regions, the 4th order moment also
begins to contribute to the form factors.
Up to about $q\approx 0.5\,\textrm{fm}^{-1}$ in $^{48}$Ca,
and $q\approx 0.35\,\textrm{fm}^{-1}$ in $^{208}$Pb, it may be possible
to explore well the 4th order moment, together with the msr.
Nevertheless, the previous papers have not focused on
the 4th order moment, as far as the authors know.
As will be shown bellow, the 4th order moments provide us with
rich information about the nuclear surface. In particular,
it is noticeable that the point neutron density, in addition to the neutron
spin-orbit density, contributes appreciably to it.

Eq.(\ref{nth}) and Eq.(\ref{4th}) give the 4th order moment
of the nuclear charge
distribution,
\begin{equation}
\avr{r^4}_c = R_{4p}+ R_{2p}+ R _{2n}
 + R_{2W}+ R_{4W_p}+ R_{4W_n}+ R_4,\label{4thm}
 \end{equation} 
 where the following abbreviations are employed
\begin{align*}
 R_{4p}&= \avr{r^4}_p,\qquad
  R_{2p}=\frac{10}{3}r_p^2\avr{r^2}_p,\qquad
  R_{2n}=\frac{10}{3}(r_+^2-r_-^2) \avr{r^2}_n\frac{N}{Z},
  \\[4pt]
  R_{2W}&= \frac{10}{3}\left\{\left(r_p^2+\frac{3}{2M^2}\right)
     \left(\avr{r^2}_{W_p}+\avr{r^2}_{W_n}
     \frac{N}{Z}\right) -\frac{r_+^2-r_-^2}{\mu_n}\avr{r^2}_{W_n}
     \frac{N}{Z}\right\}, \\[4pt]
   R_{4W_p}&= \avr{r^4}_{W_p},\qquad
    R_{4W_n}=\avr{r^4}_{W_n}\frac{N}{Z},\qquad
    R_4=\frac{5}{2}\left(r_p^4+(r_+^4-r_-^4) \frac{N}{Z}\right).
 \end{align*}
While the msr is independent of the point neutron density,
the 4th order moment depends on it through its 2nd order moment in $R_{2n}$.
Generally, the $2n$-th order moment of the charge distribution depends on
the $2(n-k)$-th\,(\,$k=1, 2\cdots,n-1$\,) order moment of the point neutron density.

Table 2 shows the 4th order moments of Eq.(\ref{4thm}) calculated
for $^{40}$Ca, $^{48}$Ca and $^{208}$Pb
by using the relativistic model NL3\cite{nl3} and NL-SH\cite{nlsh}.
As was seen in the detailed discussions on the msr, Eq.(\ref{4thm})
of the 4th order moment should be used in relativistic framework, but not
in non-relativistic calculations.
For comparison, however, we will show in Table 2 the results of the
non-relativistic calculations with SLy4\cite{sly4},
where $R_{4p}$, $R_{2p}$ and $R_{2n}$ are calculated with non-relativistic
single-particle wave functions, and the contributions of the spin-orbit
densities are obtained with use of Eq.(\ref{nsom}),
replacing its $G_\alpha$ by non-relativistic wave functions. 
The non-relativistic expression of the 4th order moment, which is equivalent
to Eq.(\ref{4thm}), may be obtained by the expansion of the nuclear
hamiltonian up to order $1/M^4$, according to the F-W transformation
\cite{tobe}.

\begin{table}
\begin{tabular}{|l||r|r|r|r|r|r|r||r|} \hline
\rule{0pt}{12pt}
 & \multicolumn{1}{c|}{$R_{4p}$} & \multicolumn{1}{c|}{$R_{2p}$} & \multicolumn{1}{c|}{$R_{2n}$}
 & \multicolumn{1}{c|}{$R_{2W}$} & \multicolumn{1}{c|}{$R_{4W_p}$} & \multicolumn{1}{c|}{$R_{4W_n}$}
 & \multicolumn{1}{c||}{$\avr{r^4}_{c}$}   &  \multicolumn{1}{c|}{Exp.} \\ \hline
\rule{0pt}{12pt}NL3    &   &  &  &  &  &  &  &  \\ 
$^{40}$Ca & $183.921$ & $ 24.941$ & $ -4.432$ & $-0.000$ & $0.457$ & $ -0.525$ & $204.952$ & $199.991$ \\
$^{48}$Ca & $178.085$ & $ 24.968$ & $ -7.278$ & $-0.283$ & $0.738$ & $ -4.962$ & $191.664$ & $194.714$ \\ 
$^{208}$Pb& $1115.64$ & $ 65.197$ & $-20.254$ & $-0.067$ & $8.269$ & $-12.304$ & $1156.81$ & $1171.58$ \\
\hline
\rule{0pt}{12pt}NL-SH      &   &  &  &  &  &   &  & \\ 
$^{40}$Ca & $178.612$ & $ 24.652$ & $ -4.382$ & $-0.000$ & $0.499$ & $ -0.569$ & $199.401$ & $199.991$ \\ 
$^{48}$Ca & $174.602$ & $ 24.805$ & $ -7.190$ & $-0.286$ & $0.780$ & $ -4.973$ & $188.134$ & $194.714$ \\
$^{208}$Pb& $1098.69$ & $ 64.839$ & $-20.046$ & $-0.068$ & $8.381$ & $-12.427$ & $1139.70$ & $1171.58$ \\
\hline
\rule{0pt}{12pt}SLy4  &    &  &  &  &  &  &   & \\
$^{40}$Ca & $194.854$ & $ 25.694$ & $ -4.555$ & $-0.000$ & $-0.073$ &  $0.061$ & $216.572$ & $199.991$ \\ 
$^{48}$Ca & $196.171$ & $ 26.126$ & $ -7.260$ & $-0.223$ & $-0.014$ & $-3.475$ & $211.722$ & $194.714$ \\
$^{208}$Pb& $1121.53$ & $ 65.155$ & $-19.371$ & $-0.051$ & $ 4.716$ & $-7.500$ & $1164.81$ & $1171.58$ \\
\hline
\end{tabular}
\caption{
The 4th order moment of the charge distribution
of $^{40}$Ca, $^{48}$Ca and $^{208}$Pb. 
The value of each term in Eq.(\ref{4thm}) is listed
in the unit of fm$^4$,
except for the one of $R_4$ which is
given by $0.590$, $0.396$ and $0.329$\,fm$^{4}$ for
$^{40}$Ca, $^{48}$Ca and $^{208}$Pb, respectively.
The experimental values are obtained by the Fourier-Bessel analyses
of data in ref.\cite{sudadata}. 
For details, see the text.
}
\end{table}

Table 2 shows that the main contribution to the value of
$\avr{r^4}_c$ comes from $R_{4p}$, as expected.
In the present calculations with the relativistic models,
more than $10\%$ correction to that
stems from the msr of the proton distribution $R_{2p}$ in $^{40}$Ca
and $^{48}$Ca, and about 6\% in $^{208}$Pb.
The sum of two terms $R_{4p}$ and $R_{2p}$, however, overestimates
the experimental values, except for the case of NL-SH for $^{208}$Pb .
In the case of SLy4 for $^{48}$Ca,
the value of $R_{4p}$ itself exceeds the experimental one.
Hence, it is necessary to have a negative contribution from the neutron
density.

In $^{48}$Ca, $R_{2n}$ reduces the value of $R_{2p}$
by about 29.1$\%$ in the case of NL3. 
The sum of  $|R_{2n}|$ and $|R_{4W_n}|$ amounts to 6.87$\%$ of $R_{4p}$,
and to 47.6$\%$ of the sum of $|R_{2p}|$ and $|R_{4W_p}|$.
Since the value of $R_{2p}$ is almost fixed by the experimental value
of the msr, we may compare $|R_{2n}+R_{4W_n}|$
with $R_{4p} - R_{2p}$ in order to see the contribution
of the neutrons to $\langle r^4 \rangle_c$. 
Then we have their ratio 7.99\%.
In NL-SH, those values are similar to, 
and in SLy4 a little smaller than the ones of NL3.

In $^{208}$Pb, the sum of  $|R_{2n}|$ and $|R_{4W_n}|$ is 44.3\%
of the sum of $|R_{2p}|$ and $|R_{4W_p}|$ in NL3.
The ratio of the sum, $|R_{2n}|$ and $|R_{4W_n}|$, to $R_{4p}$
is decreased to be 0.0292, compared with 0.0687 in $^{48}$Ca.
This is due to the constraint on the $A$-dependence of the msr in the
stable nuclei.
In more neutron-rich nuclei which are free from the constrain, 
the contribution of the neutron density to the 4th order moment
is expected to be increased appreciably. 

The parameters of the present phenomenological models\cite{nlsh, nl3, sly4}
are fixed so as for the point proton distribution to reproduce almost
the experimental values of the msr.
As a result, relativistic and non-relativistic models
yield a similar value of $R_{2p}$, as seen in Table 2.
Although, as was discussed in the previous subsection on the msr,  
$R_{4p}$ calculated in the non-relativistic model is not equivalent
to the relativistic ones, we note the following three points.
First, the values of $R_{4p}$ are different from each other by 5 to 10$\%$.
Second, in the relativistic models, $R_{4p}$ of $^{40}$Ca is
larger than the one of $^{48}$Ca,
while in SLy4 the relationship is opposite.
Third, on the one hand,
the values of the 4th order moment of the charge density
by NL3 and NL-SH are in a better agreement with experiment,
compared with the ones by SLy4 in Ca, mainly owing to the difference
between their $R_{4p}$ values.
On the other hand, in $^{208}$Pb, the relativistic models fairly
underestimate the experimental value.
These facts imply that the 4th order moment
yields valuable information as to the nuclear surface,
in addition to the one from the msr.

\section{Summary}\label{s}

The purpose of the present paper is twofold.

The first one is to make clear a role of each component of the
nuclear charge density in the mean square radius(msr).
The obtained results will be useful for refining or constructing
phenomenological nuclear models which employ the experimental values
of the msr in order to fix their parameters.
Those are also helpful for detailed analyses of experimental data
by making use of the nuclear models.

In relativistic models, the msr is dominated by the point proton density
with a few \% correction from a single-proton and -neutron size
and the neutron spin-orbit density in stable neutron-rich nuclei.

For the non-relativistic models, the expression of the msr equivalent
to the relativistic one up to order $1/M^2$ has been derived,
according to the Foldy-Wouthuysen unitary transformation.
The terms in the expression are formally consistent
with each other, but, in practical use, may not be consistent,
since non-relativistic expressions of the nucleon form factors are not
known and are usually replaced with the form factors determined by experiment.
Moreover, if parameters of non-relativistic models are fixed so as to
reproduce the experimental values of the msr without taking into account
relativistic corrections, the msr of their point proton density may contain
contribution from relativistic effects implicitly.

The second purpose is to propose a complementary method for exploring
the neutron density to the previous analyses of experiment\cite{thi}.
It is to study the $n$-th order moment, instead of the charge density
profile itself deduced from electron scattering data.
In contrast to the msr, 
the $n(\,\ge 4)$-th order moment depends not only on the neutron spin-orbit
density, but also on the point neutron density.

In the 4th order moment, a large part is explained by the point proton
density, but it is apparent that the contribution from the point
neutron density and from the neutron spin-orbit density
are necessary for reproducing the experimental values of the 4th order
moment of the charge density.

For example, in $^{48}$Ca, the main components of the 4th order moment
of the charge density are, in addition to the 4th and 2nd order moments
of the point proton density, the msr of the point neutron density and
the 4th order moments of the neutron spin-orbit density.
Among them, the 4th order moment of the point proton density dominates
the 4th order moment of the charge density,
but when 2nd moment of the point proton density is added to it,
the sum of them overestimates the experimental value.    
It is decreased by the negative contributions from the msr of
the point neutron density and the 4th order moment of
the neutron spin-orbit density.   
Their negative contributions to the 4th order moment of the point
proton density is about 7\% in $^{48}$Ca, and they almost eliminate a half of
the contribution from the 2nd order moment of the point proton density,
in using the relativistic models.

In $^{48}$Ca and $^{208}$Pb, the neutron contribution
to the 4th order moment of the charge density is limited, according to
the constraint of the $A$-dependence on the msr in stable nuclei.
In unstable neutron-rich and proton-rich nuclei,
there is not such a constraint on the point neutron density
and the neutron spin-orbit density. They are expected to play
a more important role in the 4th order moment of the charge density.
In electron scattering, the msr of the charge density reflects mainly
the point proton distribution selectively, while its 4th moment does
the point neutron density additionally.
Future experiment on unstable nuclei\cite{tsukada, tsuda}
would show not only change of the point proton distribution, but also
of the point neutron distribution as a function of $N$ or $Z$.

A role of the neutron spin-orbit density in electron scattering off
neutron rich nuclei has been investigated in more detail
in the previous paper\cite{ks}.

The present study on the 4th order moment of the
charge density may also be helpful for the analyses of the experimental
data from parity-violating electron scattering\cite{abra,cjh}. 
As was shown in Fig.1, in a region of the momentum transfer
$q = 0.475\,\textrm{fm}^{-1}$, where the experiment on $^{208}$Pb
has been performed, the 4th order moment considerably contributes
to the form factor, together with the msr.
In discussing the msr at this region, nuclear models used
for the analysis should also reproduce the 4th order moment.
In fact, the relativistic and non-relativistic models used in this
paper yield the 4th order moments which are fairly different from each other,
although they reproduce the msr in a similar way. 
The components of the 4th order moment have a clear physical meaning
separately, and hence, their detailed investigation may reduce
the ambiguity of the nuclear models.  

Finally, we note that a precise determination of the proton and
the neutron size\cite{sick} is necessary for more detailed
discussions of the moment as to the nuclear charge density.

\section*{Acknowledgment}
The authors would like to thank Professor T. Suda and Professor T. Tamae
 for useful discussions.

\end{document}